\documentclass[11pt]{article}

\usepackage[margin=1in]{geometry}

\usepackage{newtxtext,newtxmath}

\usepackage{graphicx}
\usepackage{float}
\usepackage{amsmath}
\usepackage{booktabs}
\usepackage{tabularx}
\usepackage{longtable}
\usepackage{multirow}
\usepackage{subcaption}
\usepackage[natbib=true,maxbibnames=99,style=numeric,sorting=none]{biblatex}
\usepackage{hyperref}

\makeatletter
\renewcommand{\@biblabel}[1]{#1.\hspace{1 em}}

\title{\textbf{Fiscal Dynamics in Japan under Demographic Pressure}}

\author{Goshi Aoki\\Virginia Tech\\\texttt{goshi@vt.edu}}

\date{} 

\begin{document}
\maketitle

\begin{abstract}

Japan faces a demographic transformation--a shrinking total population, a declining working-age share, and rapid elderly growth. These trends squeeze public finances from both sides--fewer people paying taxes and more people drawing on pensions and healthcare. Policy debates often focus on one fix at a time, such as raising taxes, reforming pensions, or boosting productivity. However, these levers interact with each other through feedback loops and time delays that remain insufficiently understood. Our study builds and calibrates an integrated system dynamics model that connects demographics, labor supply, economic output, and public finance to explore two questions: (RQ1) What feedback structure links demographic change to fiscal outcomes? (RQ2) Which combinations of policies can stabilize key fiscal indicators within a policy-relevant timeframe? The model, grounded in official statistics, has demonstrated its ability to track historical trends. Policy experiments show that productivity improvements and per-person controls on pension and insurance costs offer the most effective near-term relief, because they act quickly through revenue and spending channels. In contrast, raising fertility actually worsens the fiscal picture in the medium term, since it takes decades for newborns to grow up and join the workforce. A combined scenario pairing moderate productivity gains with moderate cost control nearly eliminates the deficit by 2050. These findings emphasize the importance of timing when evaluating demographic policy. Stabilizing finances within a practical timeframe requires levers that improve the budget directly, rather than those that work through slow demographic channels. The model acts as a transparent testing ground for designing time-aware fiscal policy packages in aging, high-debt economies.

\end{abstract}

\section{Introduction}
In advanced economies, increasing longevity and persistently low fertility rates are fundamentally altering public finance. Japan exemplifies this demographic transformation. A declining total population, a decreasing proportion of working-age individuals, and rapid expansion of the population aged 65 and older are simultaneously reducing the tax base and increasing age-related expenditures, thereby exerting sustained pressure on fiscal balances. Since the 1990s, Japan’s working-age population has steadily declined while the proportion of elderly individuals has risen, and official projections indicate that this trend will continue for several decades \citep{ipss2023, ipss2025}. On the expenditure side, pensions and health and long-term care now account for a substantial and increasing share of public budgets \citealp{estat2025}. This has created major concerns about Japan's ability to cover social security in the long run.

This challenge feeds on itself over time. Fiscal stress arises not only from the structural gap between revenues and age-related expenditures but also from the dynamics of debt accumulation. Each year's shortfall adds to the total debt, and a bigger debt means more interest to pay, which makes the shortfall even worse. Without timely intervention, this reinforcing cycle can accelerate.

Moreover, different solutions take different amounts of time to work. Some changes, like raising taxes or cutting benefits, can help the budget quickly. Others, like trying to boost birth rates, take decades to show results because it takes a generation for those children to grow up and start working. These delay structures help explain why experts disagree about which combination of policies—like interest rates, taxes, pension reform, or immigration—can really put Japan’s finances on a stable path soon enough.


Economists have studied Japan's aging-and-budget problem extensively using mathematical models. These models calculate how much the government would need to adjust taxes or benefits to keep its finances sustainable, and they can even estimate how different generations would be affected by various reforms \citealp{imrohoroglu2016,kitao2015,braun2016}. Other researchers have focused on how interest rates and government spending behavior can change the trajectory of Japan's debt over time \citealp{doi2011,hansen2023,hoshi2018}. While these studies provide valuable insights, they often depict policy decisions as straightforward rules or predetermined adjustment paths. In practice, policy environments are more complex; one change can trigger subsequent effects, outcomes may be delayed, and various elements are interdependent. Recognizing these interactions is crucial, as they significantly influence whether a set of reforms can achieve desired outcomes within a policy-relevant timeframe.


Given the complexity of the problem and the interconnectivity among a wide range of variables across demography and economics, a dynamic, comprehensive method that incorporates multiple factors can be insightful. During the past sixty years, system dynamics (SD) models have contributed to our understanding of complex economic problems. Studies from Jay Forrester and his World Dynamics model provided a framework to study factors that limit long term growth and lead to rapid collapse \citep{forrester1971world}. Subsequent studies, especially ones referring to as the limit to growth studies, have taken a more in-depth audit of feedback structures affecting primary and recent updates by Randers still predict major challenges for long-term growth \citep{meadows1972limits, meadows2012limits}.

There is a wide range of applications of SD to dynamic economic problems \citep{wheat2007feedback, cavana2017feedback}. Researchers have analyzed how increasing interest payments can accelerate debt accumulation, while government interventions attempt to mitigate this process \citep{arenas2003, radianti2004, burns2007, john2010, sorci2012}. 
Others have studied aging dynamics in various contexts such as health care, providing insights into relevant age-related societal problems \citep{eberlein2013}. SD models have also been applied to pension systems that finance current retirees through contributions from current workers, exploring the implications of demographic shifts on such systems \citep{petrides2004, yavas2012, stamboulis2016}.


Research applying SD to Japan has produced useful components, including detailed models of healthcare spending and long-term care insurance \citep{inoue2022, nishi2020}. Related work on Accounting System Dynamics (ASD) has built accounting frameworks and applied them to Japan’s national accounts and money flows \citep{yamaguchi2015asd, yamaguchi2022a, yamaguchi2022b}. However, comprehensive models that integrate demographics, labor supply, economic output, tax revenue, age-related expenditures, deficits, and debt within a single, data-grounded, transparent structure remain relatively scarce. This gap is significant because effective policy design for an aging, high-debt country requires understanding which chain reactions are triggered by specific policy levers. It also requires understanding how delays shift costs and benefits across time.
 

This study addresses that need by developing and calibrating an integrated SD model of Japan’s demographic--macro--fiscal system using official statistics. We focus on two research questions: (RQ1) What feedback structure links demographic change to fiscal outcomes? (RQ2) Which combinations of policies can stabilize key fiscal indicators within a policy-relevant timeframe? We use the calibrated model as a scenario laboratory, conducting policy experiments over plausible parameter ranges to compare revenue-side, spending-side, and demographic levers within a unified feedback-and-delay framework.

\section{Problem definition}

Japan’s fiscal challenge is dynamic. An aging population is shrinking the pool of taxpayers while increasing the number of people who depend on government benefits. At the same time, yearly budget shortfalls pile up into debt, and the interest on that debt can make things worse in a self-reinforcing cycle.

Since cohort transitions and fertility responses unfold with multi-decade delays, policies that help in the long run—such as encouraging higher birth rates—can, in fact,  make the budget look worse in the short run. Furthermore, the longer the government waits to act, the larger the adjustment it will eventually need.

\begin{figure}[H]
\centering
\begin{minipage}{0.499\linewidth}\centering
\includegraphics[width=\linewidth]{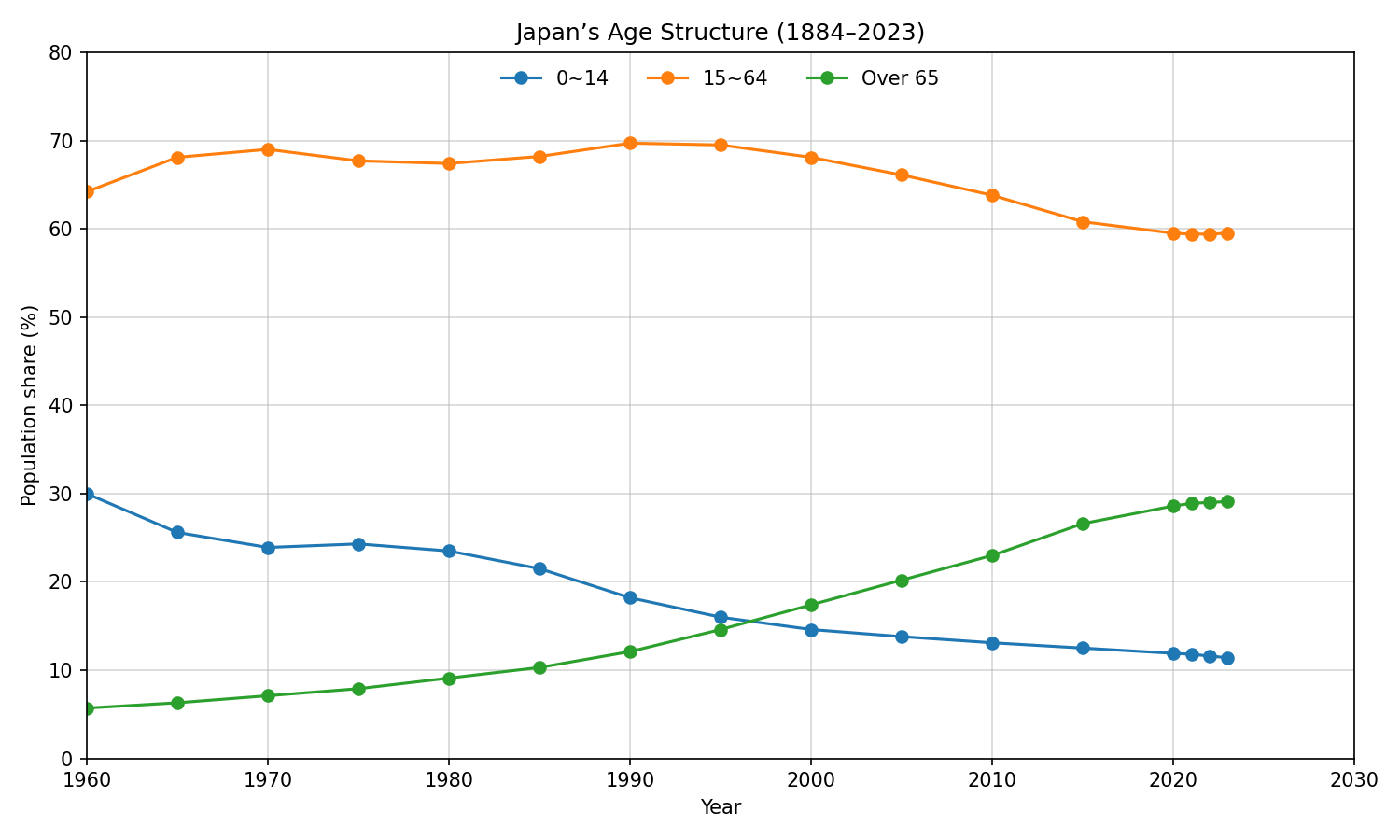}
\subcaption{Japan's age structure (declining 15--64 share and rising 65+ share).}
\label{fig:age_structure}
\end{minipage}\hfill
\begin{minipage}{0.499\linewidth}\centering
\includegraphics[width=\linewidth]{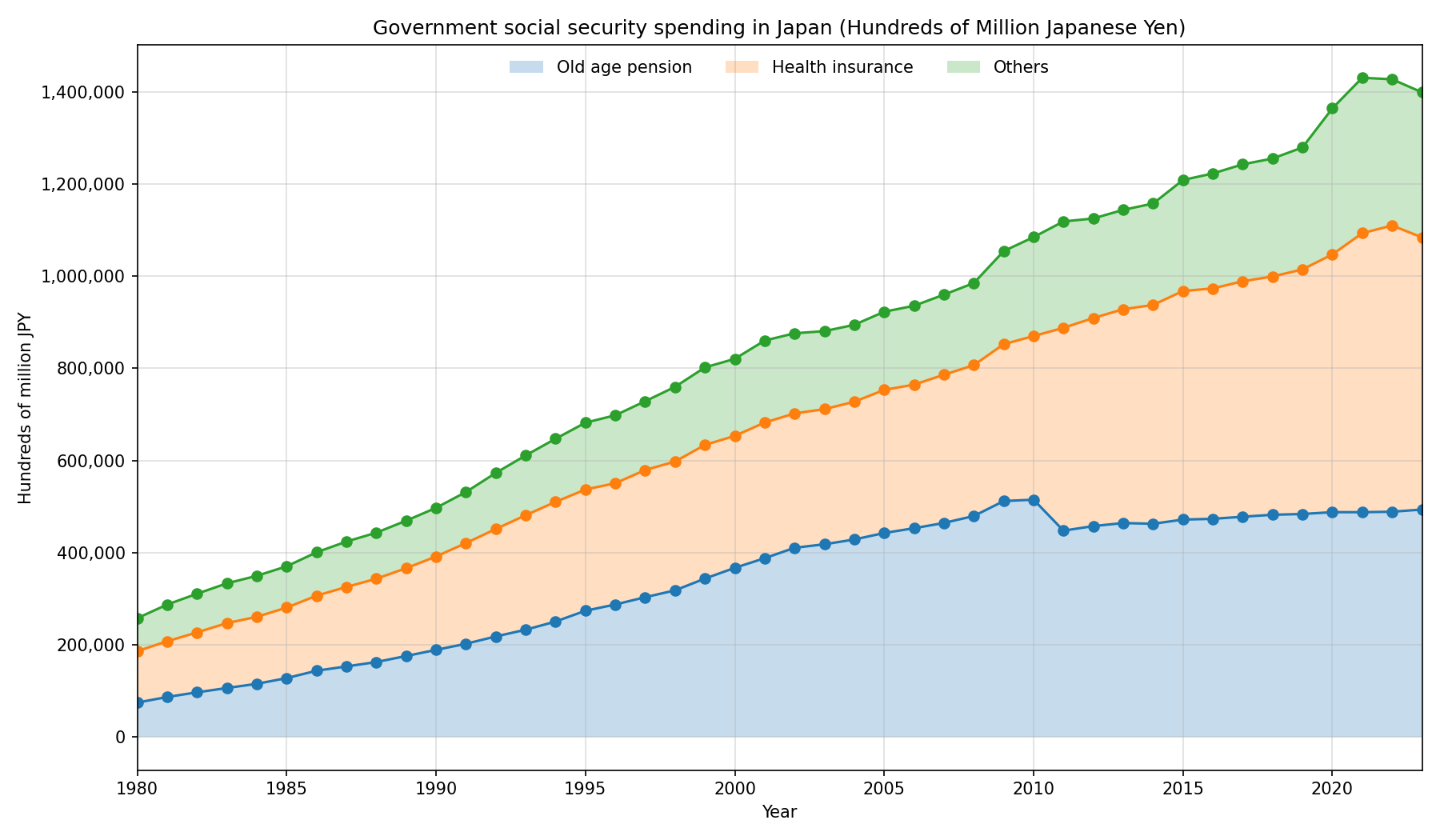}
\subcaption{Social-security spending in Japan.}
\label{fig:social_security}
\end{minipage}
\caption{Reference modes for Japan's demographic and fiscal trends.}
\label{fig:ref_modes_1}
\end{figure}

We summarize the problem using reference modes that motivate the model boundary and feedback structure.
Figure~\ref{fig:ref_modes_1} shows (a) the age-structure shift—working-age share peaking near $\sim$70\% in the 1990s and trending downward while the 65+ share rises (IPSS; \citealp{ipss2025})—and (b) the rise in social-security spending (e-Stat; \citealp{estat2025}).

\begin{figure}[H]
\centering
\begin{minipage}{0.49\linewidth}\centering
\includegraphics[width=\linewidth]{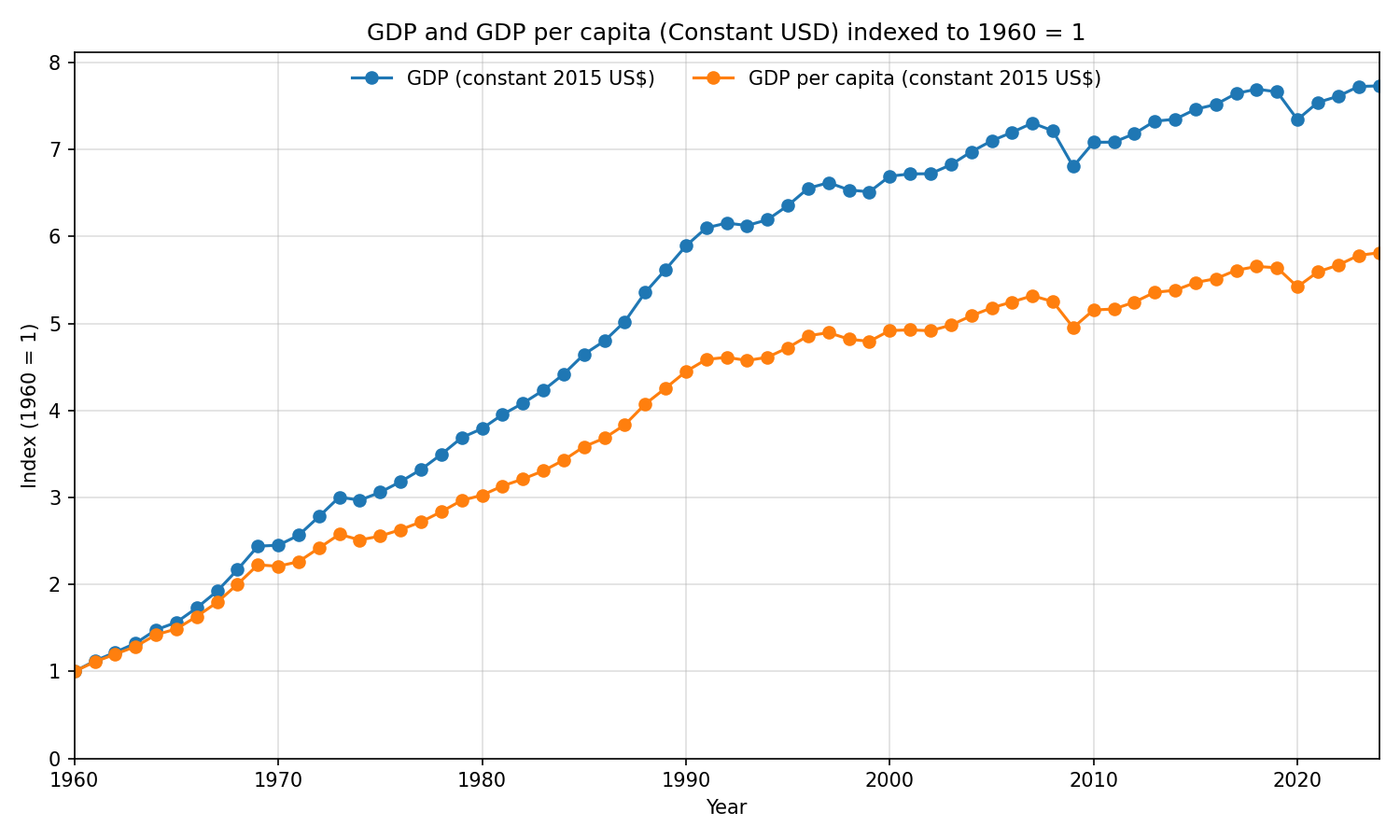}
\subcaption{GDP and GDP per capita after $\sim$1990 (indexed).}
\label{fig:gdp_per_capita}
\end{minipage}\hfill
\begin{minipage}{0.49\linewidth}\centering
\includegraphics[width=\linewidth]{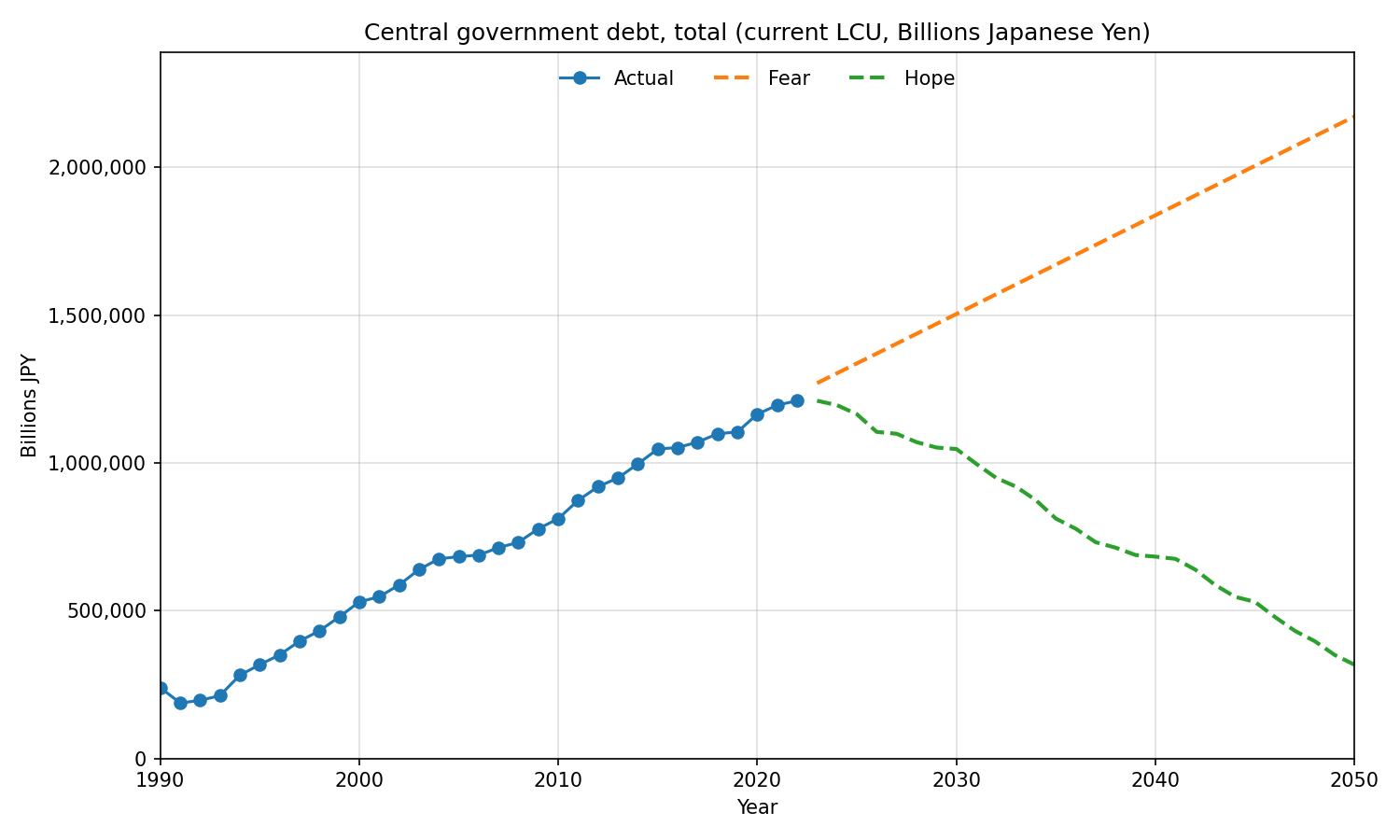}
\subcaption{Government debt with divergent future trajectories.}
\label{fig:gov_debt}
\end{minipage}
\caption{Reference modes for Japan's economic and fiscal outlook.}
\label{fig:ref_modes_2}
\end{figure}

Figure~\ref{fig:ref_modes_2} shows (a) slower post-1990 growth in GDP and GDP per capita (World Bank Open Data \citealp{wb_data_gdp, wb_data_gdp_per_capita}) and (b) the strong rise in government debt since the 2000s (World Bank Open Data; \citealp{wb_data_debt}).
Over the same period, Japan’s tax and social-security burden ratio has trended upward, indicating increasing reliance on taxes and contributions (Ministry of Finance, Japan; \citealp{mof_burden}).


These patterns motivate a dynamic hypothesis centered on competing feedbacks.
A reinforcing loop links the debt stock to interest costs and back to debt, while balancing mechanisms—especially fiscal reaction through taxation and labor-supply responses—can stabilize deficits.

\section{Method}
We follow standard system dynamics practice: problem articulation, stock–flow model formulation, empirical and structural testing, and policy analysis. 

The model integrates (i) cohort‑based demographics, (ii) labor supply and productivity‑driven output, and (iii) a fiscal accounting sector in which revenues and age‑related expenditures determine the deficit that accumulates into net debt. It is implemented in Vensim for 1968–2050 with a 0.25‑year time step.

Reference modes and calibration targets are constructed from official Japanese statistics and widely used international datasets (Appendix A). Parameters that are directly observable are set from data; remaining behavioral and institutional parameters are calibrated by minimizing deviations between simulated trajectories and historical series over their available sample periods. 

Model credibility is assessed by (i) visual comparisons to historical reference modes and (ii) quantitative fit statistics ($R^2$, MAPE, Theil's inequality statistic). 

We then evaluate three policy levers introduced from 2025—productivity growth, higher fertility, and per‑capita cost containment—each in moderate and aggressive variants, plus a combined scenario, comparing outcomes for the deficit and net debt.

\section{Structure}

\subsection{Boundary and purpose}

The model integrates three sectors: (i) demographics (population by age), (ii) labor supply and output (workforce, productivity, GDP), and (iii) public finance (revenue, spending, deficit, debt).
Its purpose is to identify the feedback mechanisms that generate fiscal pressure and to evaluate candidate policy levers.

Table~\ref{tab:model_boundary} summarizes endogenous, exogenous, and excluded variables.

\begin{table}[H]
\centering
\caption{Model boundary chart.}
\label{tab:model_boundary}
\begin{tabularx}{\linewidth}{>{\raggedright\arraybackslash}X >{\raggedright\arraybackslash}X >{\raggedright\arraybackslash}X}
\toprule
\textbf{Endogenous} & \textbf{Exogenous (baseline drivers)} & \textbf{Excluded (in this version)}\\
\midrule
Population by cohort; births/deaths & Fertility trend (baseline); mortality trend & Net immigration\\
Workforce; participation & Interest rate (baseline) & Monetary policy; exchange rates\\
GDP & Productivity trend (baseline) & Asset prices; inflation dynamics\\
Revenue; spending; deficit; debt & Policy parameters for experiments & ---\\
Tax rate (reactive rule) & --- &  \\
\bottomrule
\end{tabularx}
\end{table}

\clearpage
\subsection{Dynamic hypothesis: feedback structure}

The model's behavior is the result of the interaction of several feedback loops, which are introduced below and range from demographics to debt.

\begin{figure}[H]
\centering
\includegraphics[width=0.90\linewidth]{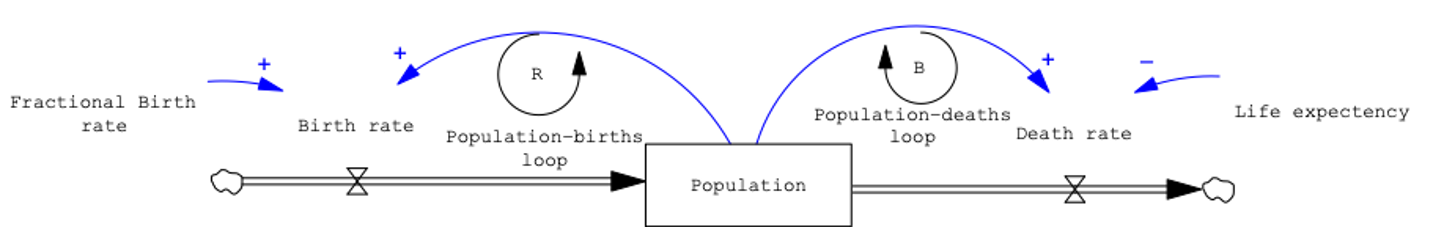}
\caption{Population dynamics: reinforcing births (R1) and balancing deaths (B1).}
\label{fig:cld-population}
\end{figure}

\textbf{Population dynamics (R1, B1).} 
Population is a stock that increases through births and decreases through deaths (Figure~\ref{fig:cld-population}). 
The \textbf{Population-births loop (R1)} captures demographic momentum: larger childbearing cohorts generate more births. 
The \textbf{Population-deaths loop (B1)} provides balancing mortality pressure.

\begin{figure}[H]
\centering
\includegraphics[width=0.80\linewidth]{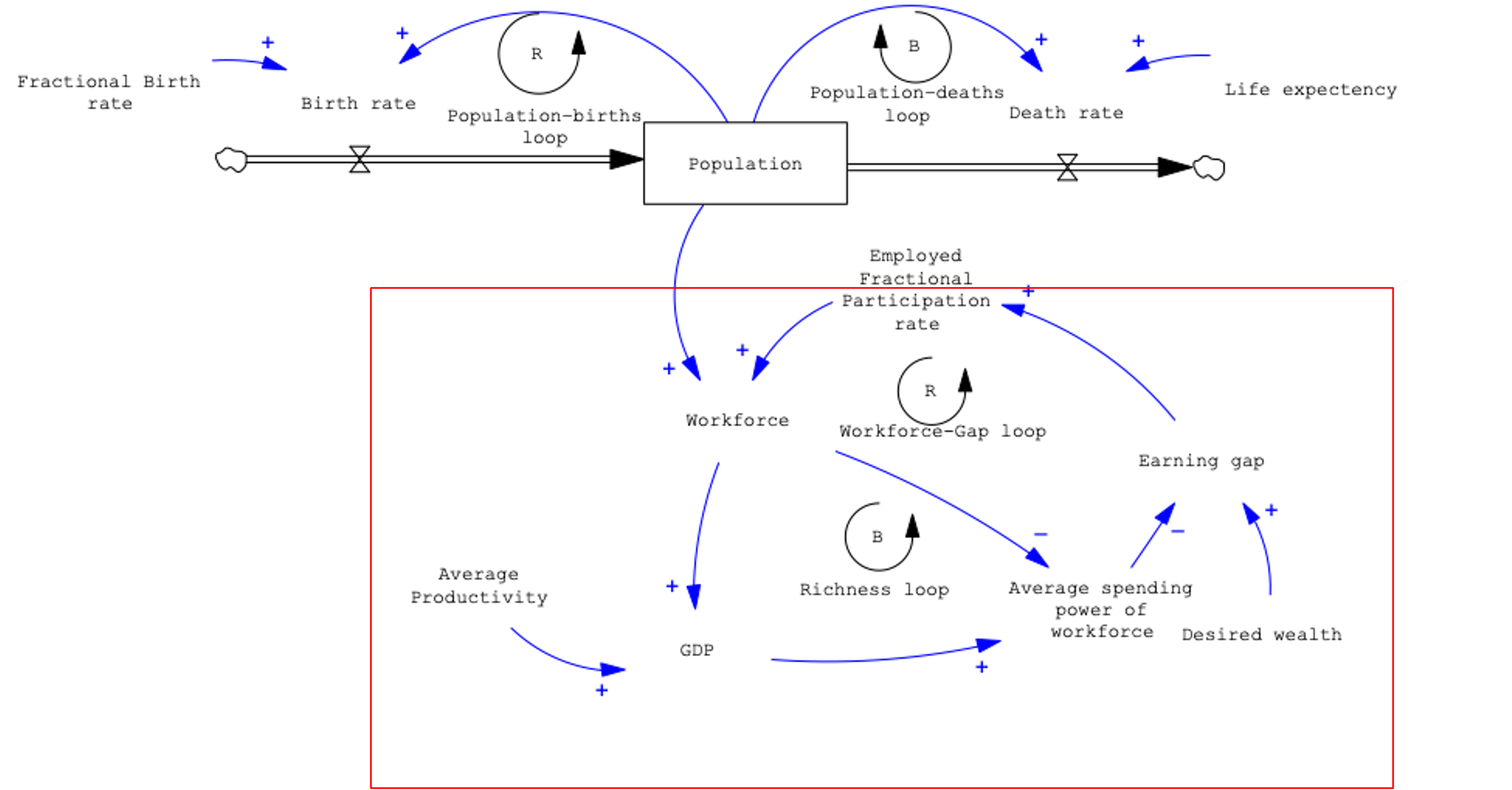}
\caption{From population to workforce and GDP: participation response and productivity.}
\label{fig:cld-workforce-gdp}
\end{figure}

\textbf{Workforce and output (R2, B2).} 
Population determines workforce through participation behavior (Figure~\ref{fig:cld-workforce-gdp}); workforce combined with average productivity produces GDP. 
The \textbf{Workforce-Gap loop (R2)} captures participation response when an earning gap encourages labor supply. 
The \textbf{Richness loop (B2)} works in the opposite direction: as GDP rises, spending power improves and the incentive to work weakens.

\begin{figure}[H]
\centering
\includegraphics[width=0.75\linewidth]{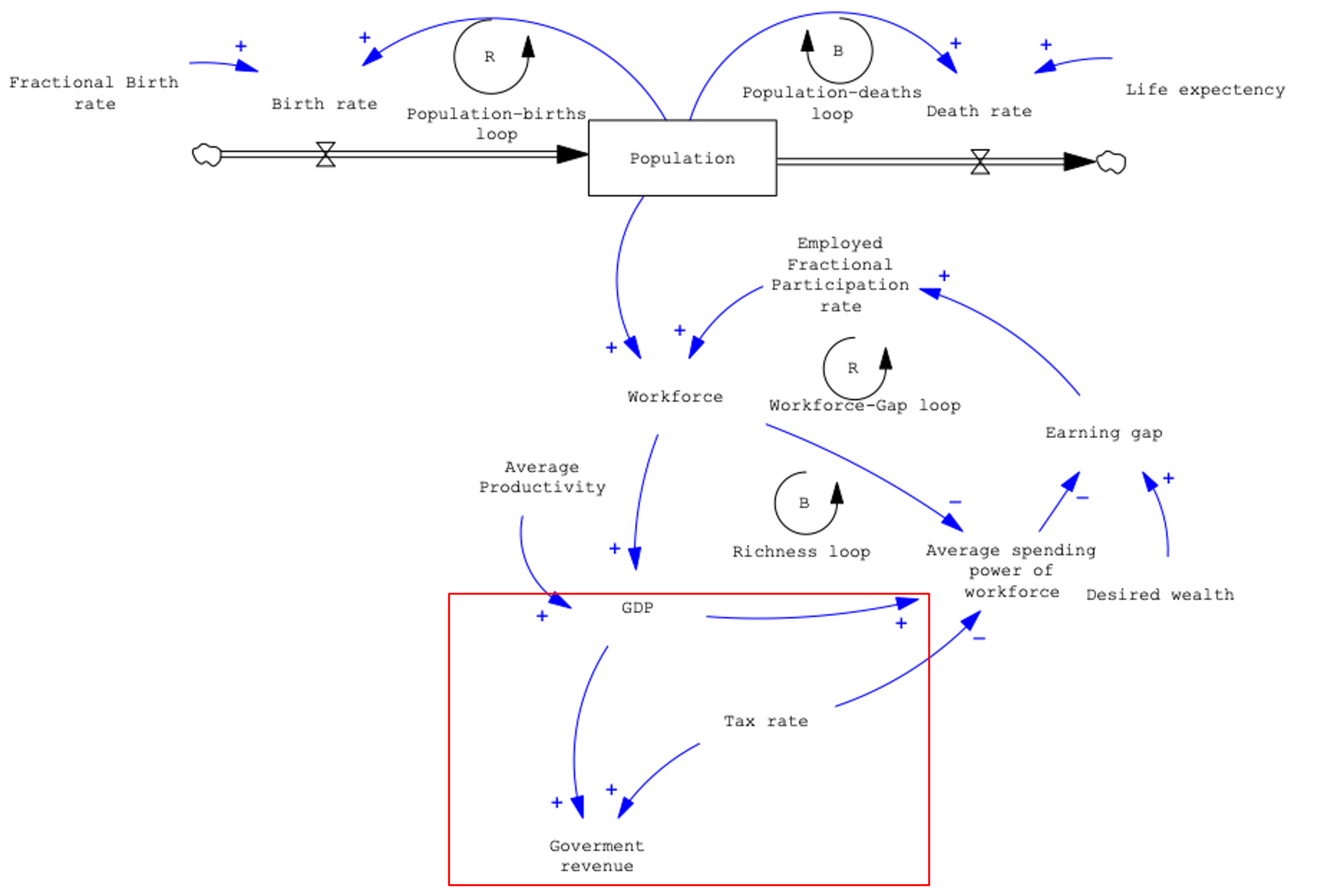}
\caption{Revenue structure: tax base and tax rate determine revenue.}
\label{fig:cld-revenue}
\end{figure}

\clearpage
\textbf{Revenue.} 
Government revenue is modeled as \textit{Tax rate} $\times$ \textit{GDP} plus other revenue (Figure~\ref{fig:cld-revenue}). 

\begin{figure}[H]
\centering
\includegraphics[width=0.8\linewidth]{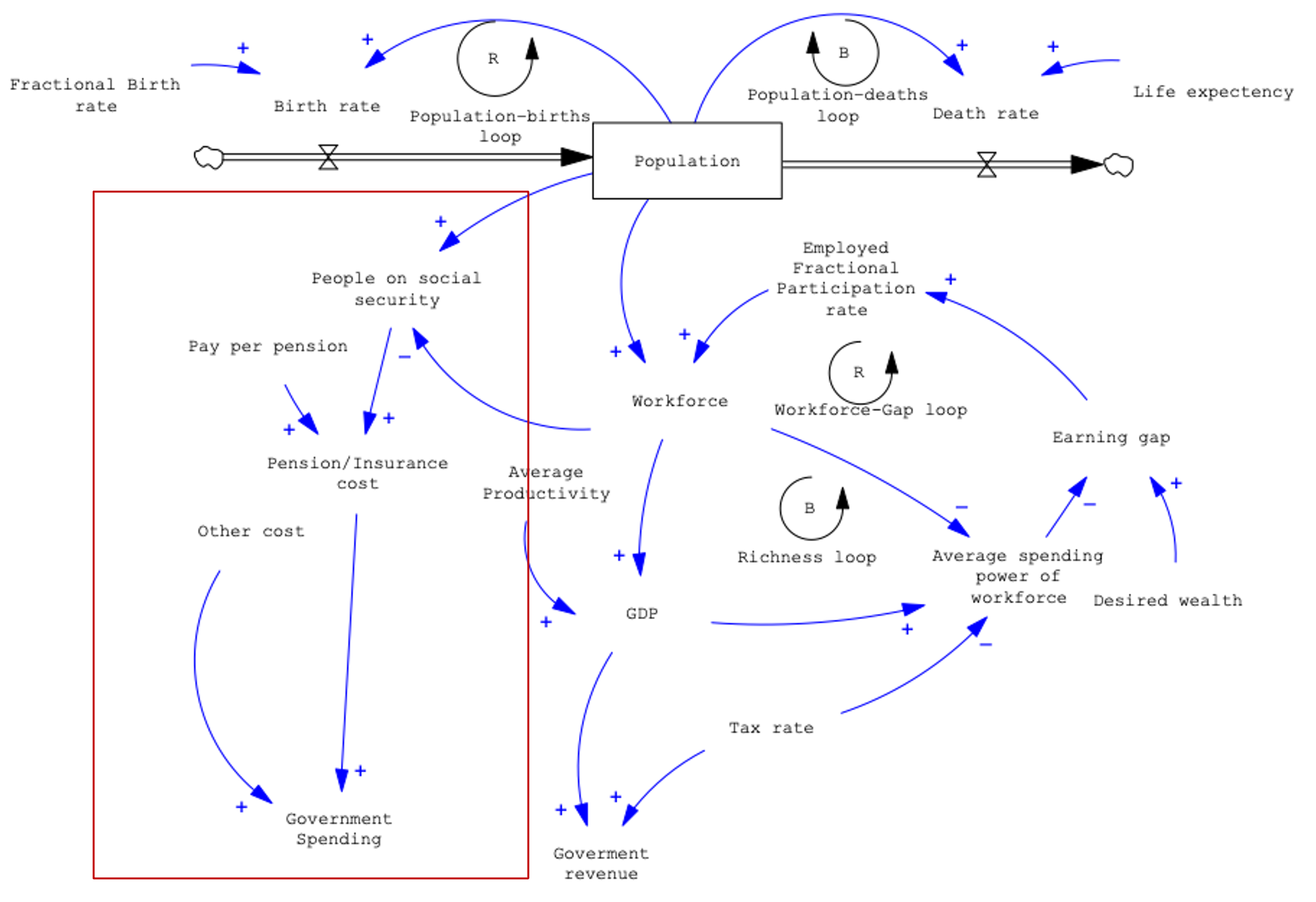}
\caption{Spending structure: beneficiaries and per-capita cost levels drive spending.}
\label{fig:cld-spending}
\end{figure}

\textbf{Age-driven spending.} 
As population ages, the number of people receiving social-security benefits rises (Figure~\ref{fig:cld-spending}). 
Multiplying beneficiaries by per-capita benefit/cost levels produces pension/insurance spending. Also, other government costs are added separately.

\begin{figure}[H]
\centering
\includegraphics[width=1\linewidth]{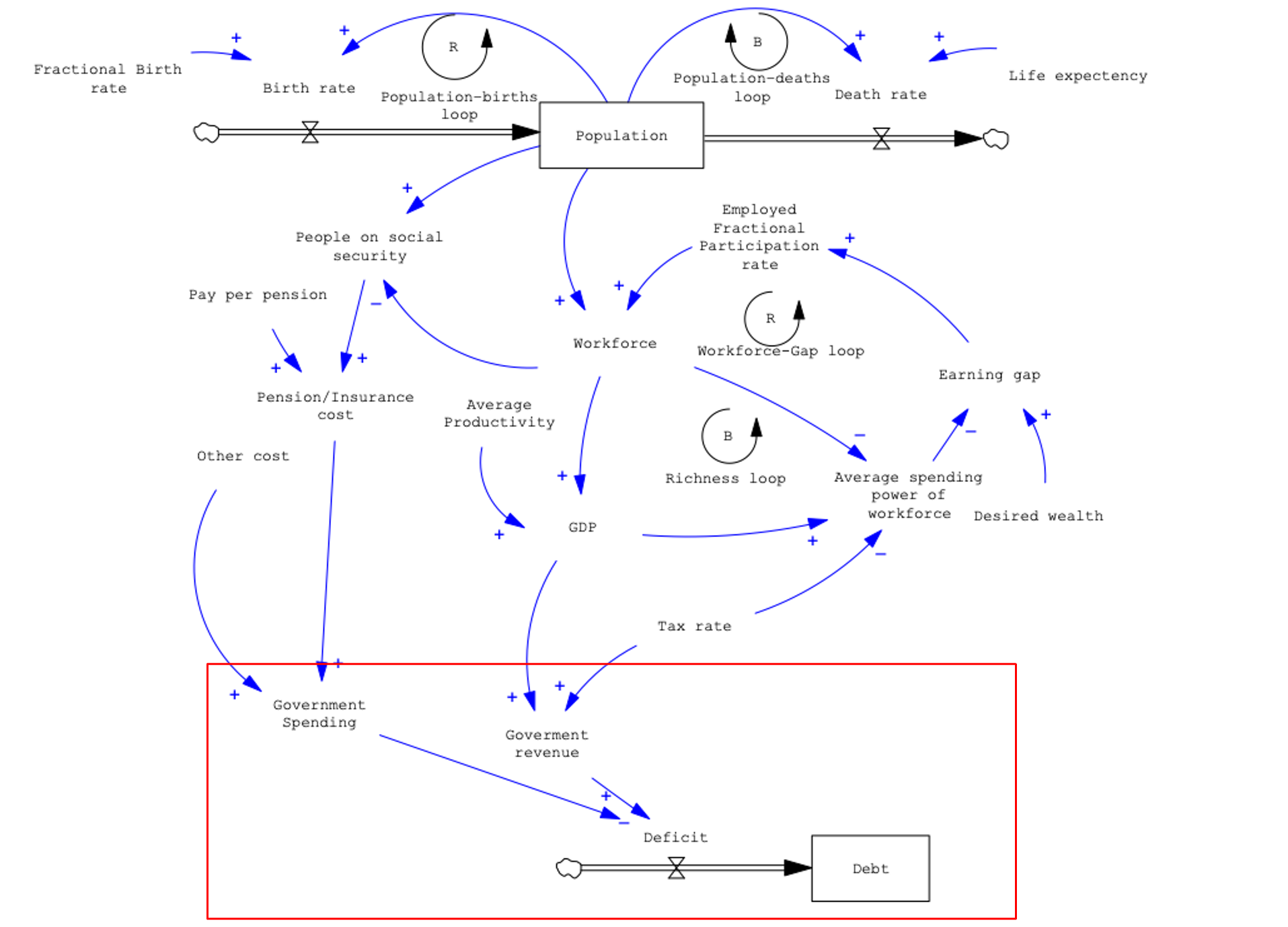}
\caption{Debt stock as accumulation of deficits.}
\label{fig:cld-debt-stock}
\end{figure}

\textbf{Debt accumulation.} 
Deficit (spending minus revenue) accumulates into a debt stock (Figure~\ref{fig:cld-debt-stock})
\begin{equation}
\frac{dD}{dt}=\text{Deficit}=\text{Spending}-\text{Revenue}.
\end{equation}

\begin{figure}[H]
\centering
\includegraphics[width=1\linewidth]{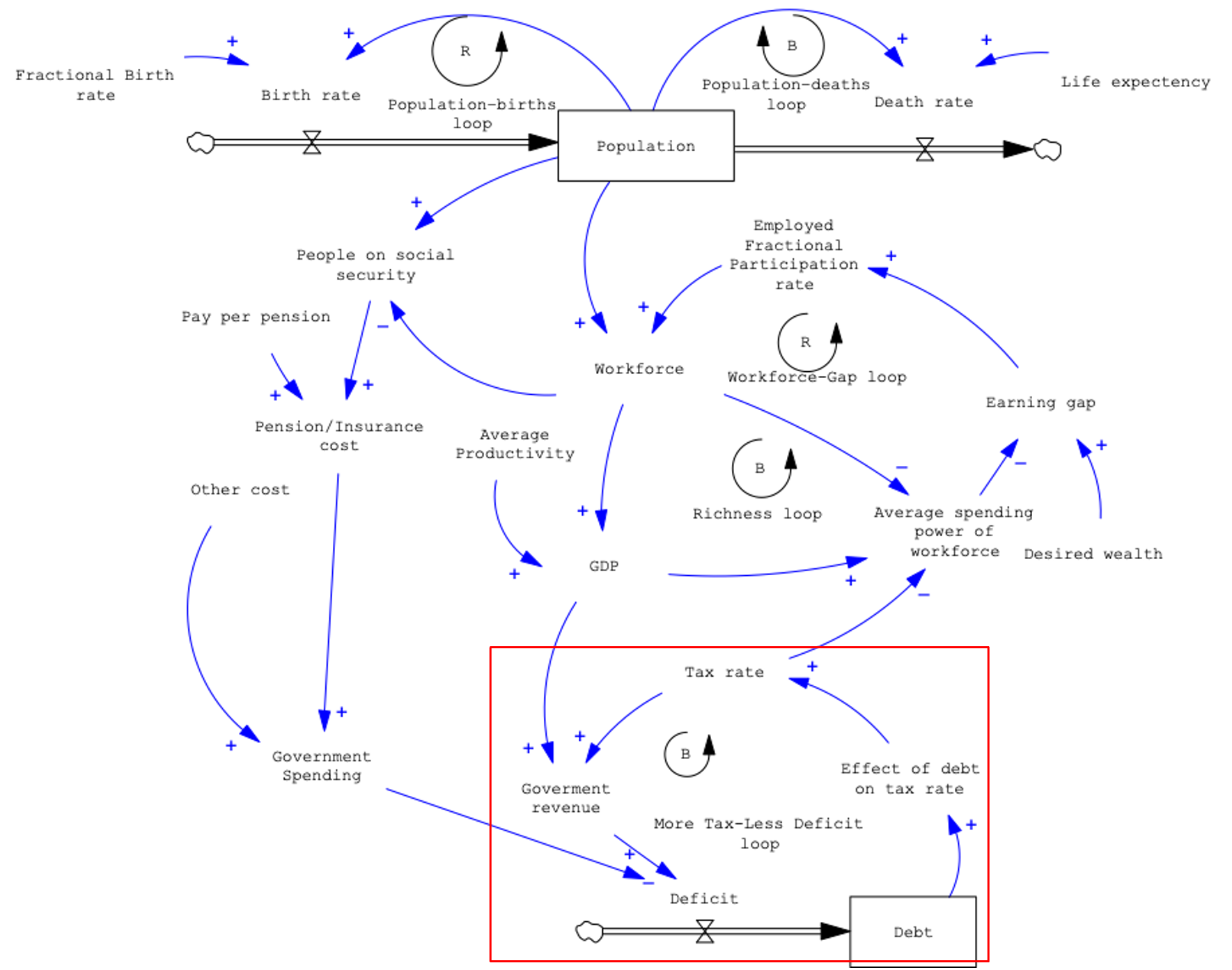}
\caption{Tax adjustment balancing loop (B3).}
\label{fig:cld-tax-stabilizer}
\end{figure}

\clearpage
\textbf{Tax stabilizer (B3).} 
A fiscal reaction rule raises the tax rate in response to rising debt/deficit pressure, increasing revenue and balancing the system (Figure~\ref{fig:cld-tax-stabilizer}). 
Implementation delays represent legislative and economic adjustment lags.

\begin{figure}[H]
\centering
\includegraphics[width=1\linewidth]{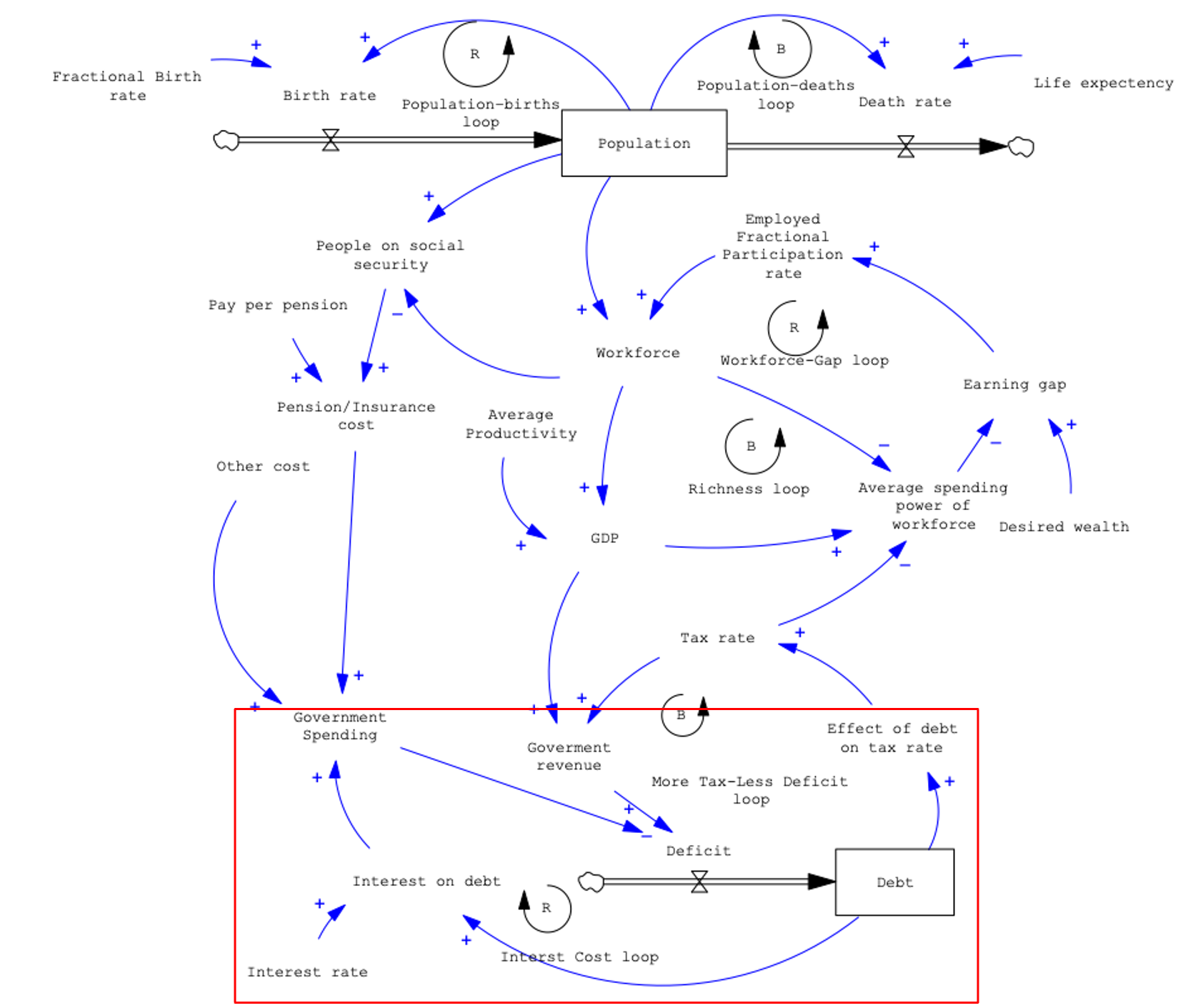}
\caption{Interest-cost reinforcing loop (R3).}
\label{fig:cld-interest-loop}
\end{figure}

\clearpage
\textbf{Interest-cost snowball (R3).} 
Interest payments increase with the debt stock:
\begin{equation}
\text{Interest} = r \times D .
\end{equation}
Higher debt raises interest payments, increasing spending, widening deficits, and further increasing debt (Figure~\ref{fig:cld-interest-loop}). 

\begin{figure}[H]
\centering
\includegraphics[width=1\linewidth]{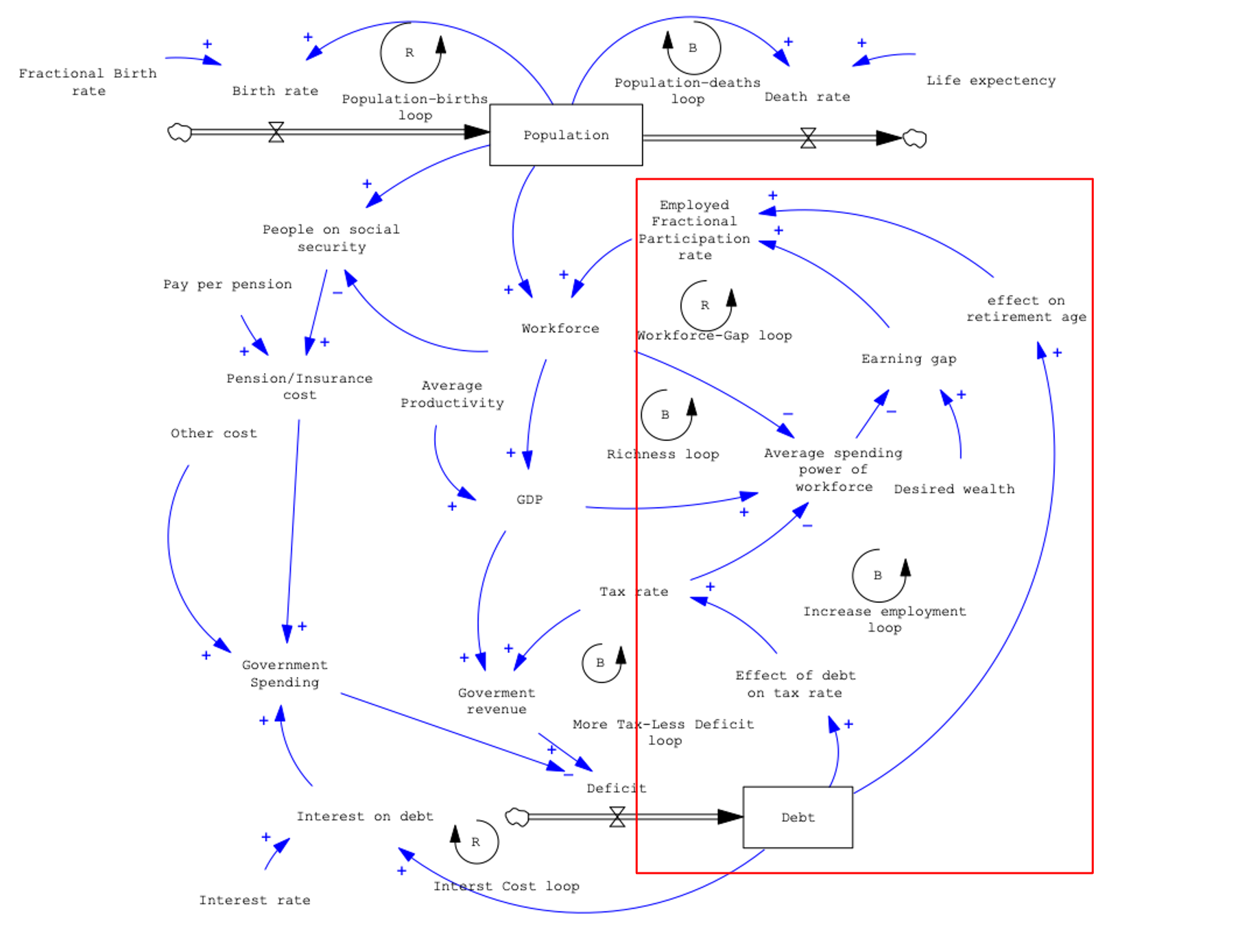}
\caption{Employment and retirement balancing loop (B4).}
\label{fig:cld-employment}
\end{figure}

\clearpage
\textbf{Employment response (B4).} 
Higher participation and later retirement increase the workforce, raising GDP and revenue and reducing deficits (Figure~\ref{fig:cld-employment}). 

\subsection{Integrated Stock--Flow Structure}
\begin{figure}[p]
\centering
\includegraphics[width=0.95\textheight, angle=90]{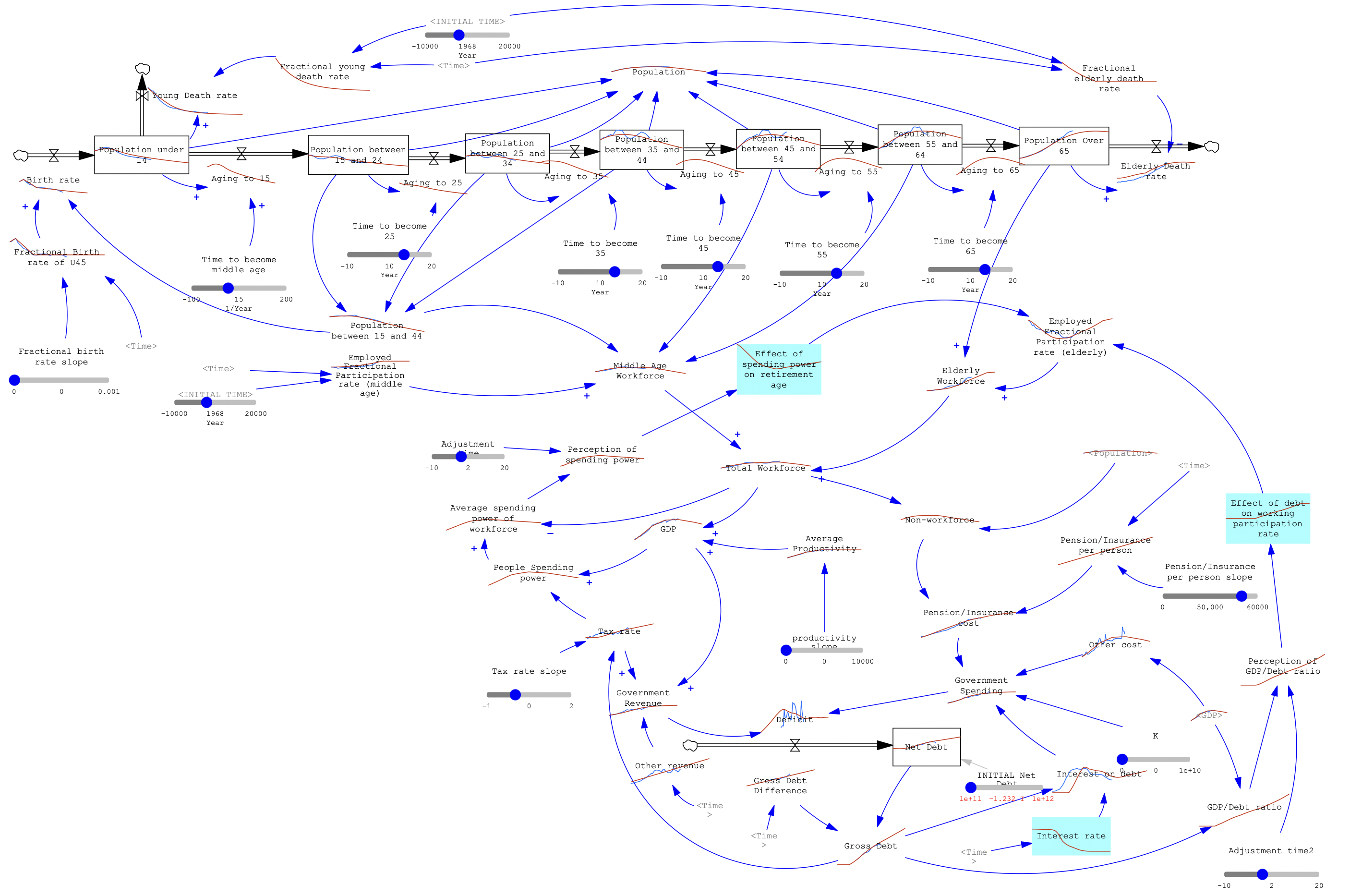}
\caption{Integrated simulation model: aging chain and debt accumulation structure.}
\label{fig:model_overview}
\end{figure}

Figure~\ref{fig:model_overview} summarizes the integrated stock--flow structure that links demographics, the macroeconomy, and public finance. 
A seven-cohort aging chain determines workforce availability; GDP is generated by labor input and labor productivity; government revenue and spending determine the deficit, which accumulates into a net-debt stock. 
The feedback loops identified in the reference modes are represented explicitly as causal links and decision rules. 

\section{Simulation result}

This section reports the main simulation outputs. 
We first show how the model reproduces key historical reference modes and summarize quantitative fit statistics. 
We then report baseline behavior and policy experiments using outcome measures for deficit and net debt. The structural explanations behind these results are discussed in the next section.

\subsection{Baseline behavior and historical fit: population and GDP}

Total population is the sum of seven age-cohort stocks:
\begin{align}
\text{POP} &= P_{0\text{--}14}+P_{15\text{--}24}+P_{25\text{--}34}+P_{35\text{--}44}+P_{45\text{--}54}+P_{55\text{--}64}+P_{65+}.
\end{align}

Aging flows between these stocks implement a continuous-cohort approximation (material delays) commonly used for representing aging populations in SD models \citep{eberlein2013}.

GDP links labor input to productivity:
\begin{align}
\text{GDP} &= \text{Productivity} \times \text{Workforce}.
\end{align}

\begin{figure}[H]
\centering
\includegraphics[width=0.6\linewidth]{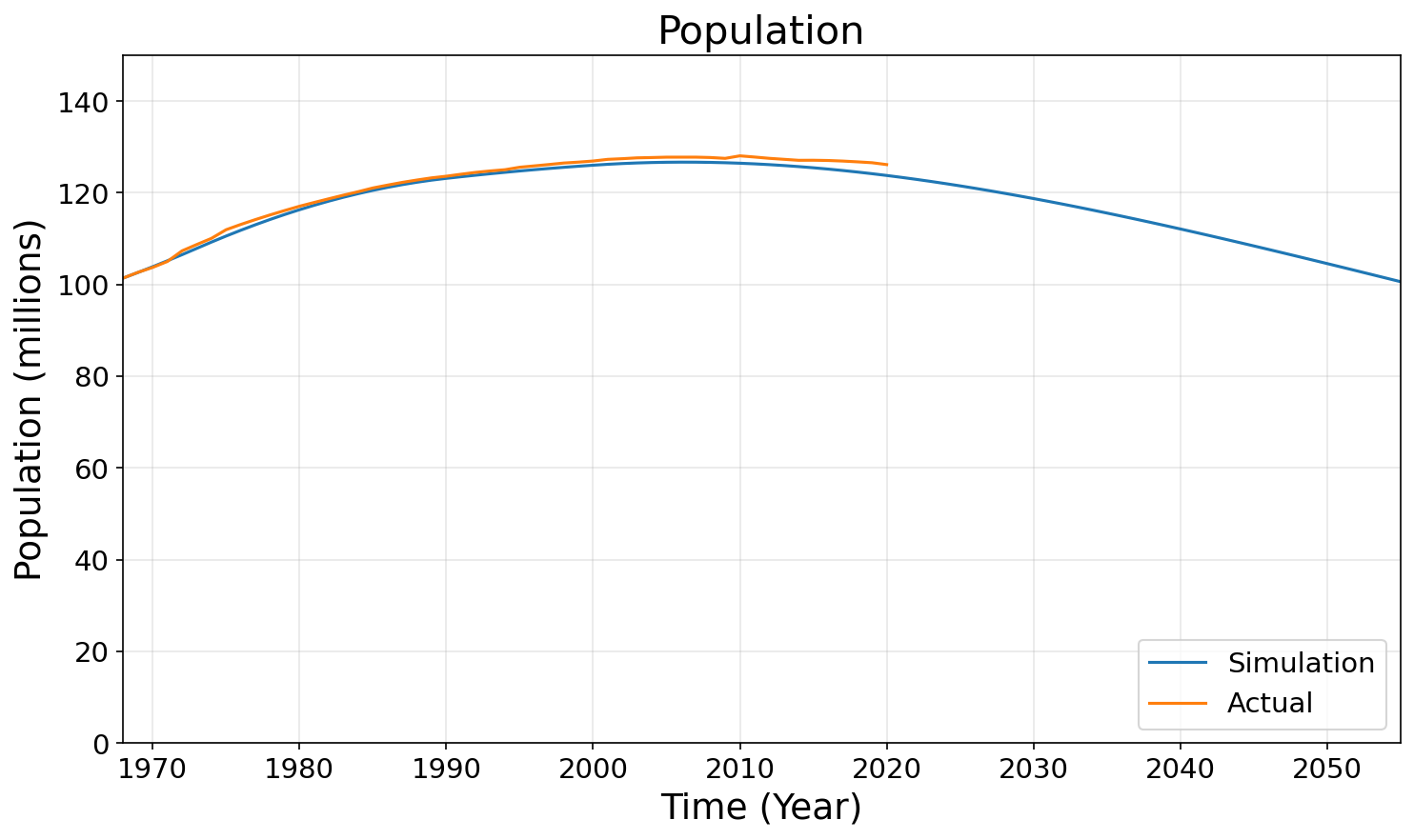}
\caption{Historical fit: simulated vs.\ observed total population.}
\label{fig:test_population}
\end{figure}

\begin{figure}[H]
\centering
\includegraphics[width=0.6\linewidth]{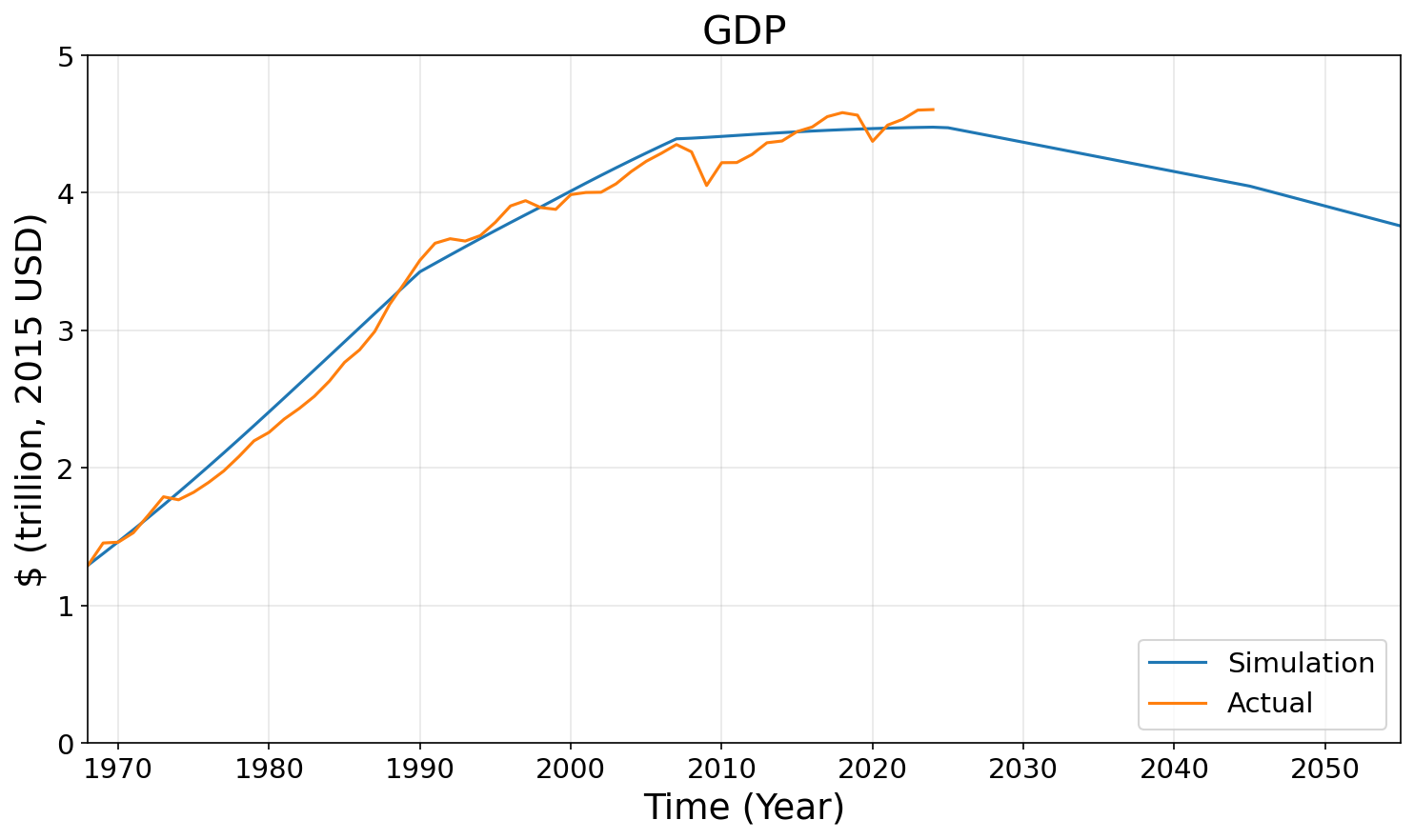}
\caption{Historical fit: simulated vs.\ observed GDP.}
\label{fig:test_gdp}
\end{figure}

Figures~\ref{fig:test_population} and \ref{fig:test_gdp} compare simulated trajectories to historical data. The model is intended to capture long-run trends rather than business-cycle fluctuations.

\subsection{Baseline behavior and historical fit}

\begin{figure}[H]
\centering
\includegraphics[width=0.9\linewidth]{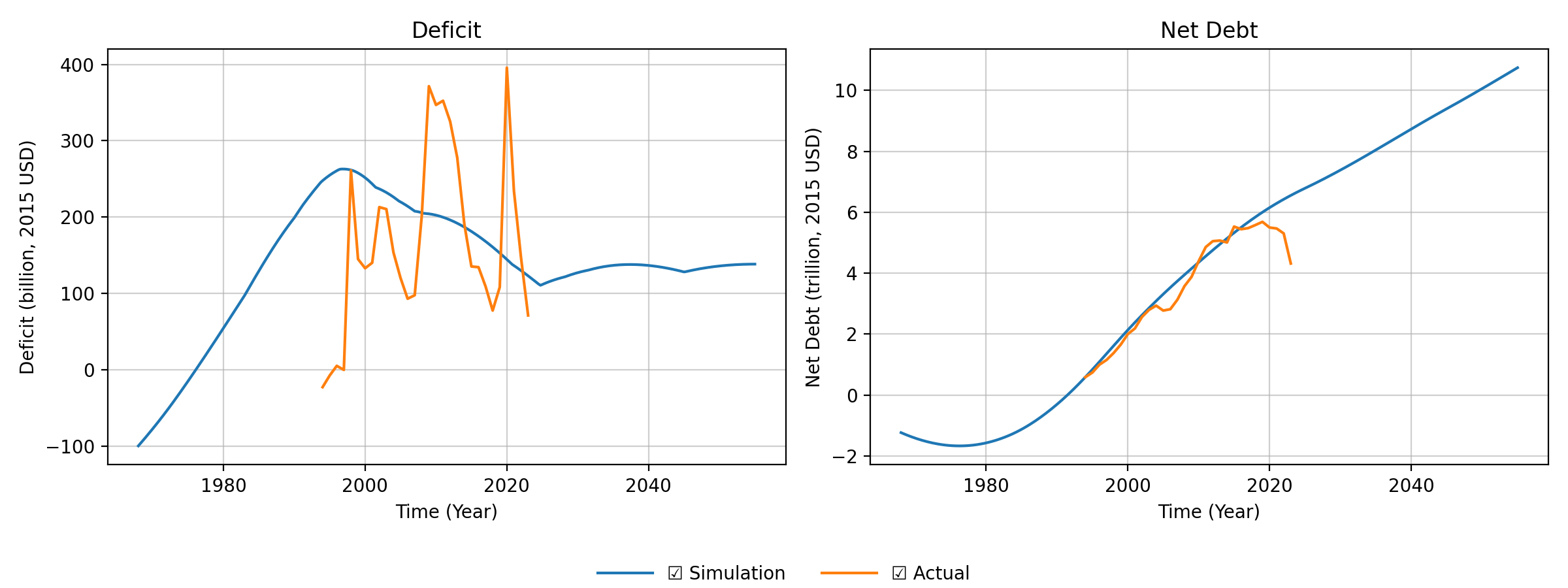}
\caption{Historical fit: deficit and net debt behavior.}
\label{fig:test_fiscal}
\end{figure}

Figure~\ref{fig:test_fiscal} compares simulated fiscal behavior to historical deficit and net debt series. Monetary quantities are reported in \textbf{trillion constant 2015 USD}. Deficit is defined as \emph{spending minus revenue}, so positive values indicate a deficit (negative indicates a surplus). For intuition,\footnote{Using the World Bank WDI official exchange rate for Japan in 2015 (LCU per USD), 1 trillion 2015 USD \(\approx\) 121 trillion 2015 JPY. \citep{wb_data_gdp}} we also report approximate conversions to ``trillion JPY (2015)'' in summary tables.

The simulation captures the persistent structural deficit that Japan has maintained since the early 2000s, driven by chronic imbalances between government spending and revenue. However, the actual deficit series exhibits pronounced short-term fluctuations that the model does not reproduce. These swings are largely attributable to discrete political decisions in response to major economic shocks---most notably the large-scale stimulus packages enacted during the Global Financial Crisis of 2008--2009 and the COVID-19 pandemic, which caused sharp upward spikes in the deficit, followed by periods of partial fiscal consolidation that temporarily narrowed it. Because the model represents underlying demographic and structural fiscal dynamics rather than exogenous policy interventions, it smooths over these event-driven fluctuations and instead tracks the long-run deficit trend. The net debt series, which accumulates these deficits over time and is therefore less sensitive to year-to-year volatility, shows a closer fit between simulation and historical data.

\subsection{Quantitative fit statistics}

Table~\ref{tab:validation_stats} summarizes how closely the model tracks historical data. We report $R^2$, mean absolute percentage error (MAPE) and Theil inequality statistics over the available historical periods (population: 1968--2020; GDP: 1968--2024; net debt: 1994--2023). 

\begin{table}[H]
\centering
\caption{Fit statistics against historical data (population: 1968--2020; GDP: 1968--2024; net debt: 1994--2023). Theil $U$ decomposition reports bias ($U^M$), variance ($U^S$), and covariance ($U^C$) proportions (summing to 1).}
\label{tab:validation_stats}
\small
\begin{tabular}{lcccccc}
\toprule
\textbf{Variable} & \textbf{$R^2$} & \textbf{MAPE (\%)} & \textbf{Theil $U$} & \textbf{$U^M$ (bias)} & \textbf{$U^S$ (variance)} & \textbf{$U^C$ (covariance)}\\
\midrule
Total population & 0.995 & 0.8 & 0.005 & 0.74 & 0.07 & 0.19\\
GDP & 0.989 & 3.1 & 0.016 & 0.12 & 0.02 & 0.85\\
Net debt & 0.931 & 10.5 & 0.068 & 0.27 & 0.05 & 0.68\\
\bottomrule
\end{tabular}
\end{table}

With the model's historical performance established, we turn to policy experiments. Three levers are tested, each targeting a different feedback channel: productivity growth (revenue side), higher fertility (demographic base), and per-capita cost containment (spending side). We also test a combined scenario.

\subsection{Scenario design}
All policy changes begin in 2025 and ramp up linearly toward 2050 targets. Table~\ref{tab:policy_summary} summarizes the design and 2050 outcomes. Productivity is increased by +10\% (moderate) or +30\% (aggressive) above the post-2007 baseline level by 2050. The birth-rate parameter is raised by +15\% or +50\% by 2050. Per-capita pension/insurance cost is capped at 10\% or 25\% below the baseline trend by 2050.

\begin{table}[H]
\centering
\caption{Policy experiments (introduced in 2025) and 2050 outcomes. Monetary values are in trillion constant 2015 USD; in parentheses we report approximate trillion JPY (2015) conversions and percent changes relative to the baseline 2050 value.}
\label{tab:policy_summary}
\small
\begin{tabularx}{\linewidth}{>{\raggedright\arraybackslash}p{0.18\linewidth} >{\raggedright\arraybackslash}p{0.32\linewidth} >{\raggedright\arraybackslash}X >{\raggedright\arraybackslash}X}
\toprule
\textbf{Lever} & \textbf{Scenario (2050 target; introduced 2025)} & \textbf{Deficit in 2050} & \textbf{Net debt in 2050}\\
\midrule
Baseline & No policy change & 0.136~trn USD (16.5~trn JPY) & 10.057~trn USD (1217~trn JPY)\\
\midrule
\multirow{2}{*}{Productivity} & Moderate: +10\% productivity by 2050 & 0.111~trn USD (-17.9\%); (13.5~trn JPY) & 9.783~trn USD (-2.7\%); (1184~trn JPY)\\
& Aggressive: +30\% productivity by 2050 & 0.095~trn USD (-30.1\%); (11.5~trn JPY) & 9.348~trn USD (-7.1\%); (1131~trn JPY)\\
\midrule
\multirow{2}{*}{Birth rate} & Moderate: +15\% birth-rate parameter by 2050 & 0.149~trn USD (+10.0\%); (18.1~trn JPY) & 10.195~trn USD (+1.4\%); (1234~trn JPY)\\
& Aggressive: +50\% birth-rate parameter by 2050 & 0.182~trn USD (+34.1\%); (22.0~trn JPY) & 10.520~trn USD (+4.6\%); (1273~trn JPY)\\
\midrule
\multirow{2}{*}{Cost containment} & Moderate: cap per-capita cost 10\% below baseline trend by 2050 & 0.031~trn USD (-77.4\%); (3.7~trn JPY) & 9.598~trn USD (-4.6\%); (1162~trn JPY)\\
& Aggressive: cap per-capita cost 25\% below baseline trend by 2050 & -0.065~trn USD (surplus); (-7.9~trn JPY) & 7.615~trn USD (-24.3\%); (922~trn JPY)\\
\midrule
Combined & Moderate bundle: +10\% productivity and 10\% cost cap & 0.005~trn USD (-96.4\%); (0.6~trn JPY) & 9.320~trn USD (-7.3\%); (1128~trn JPY)\\
\bottomrule
\end{tabularx}
\end{table}

\clearpage
\subsection{Scenario 1: Productivity growth}

\begin{figure}[H]
\centering
\includegraphics[width=0.55\linewidth]{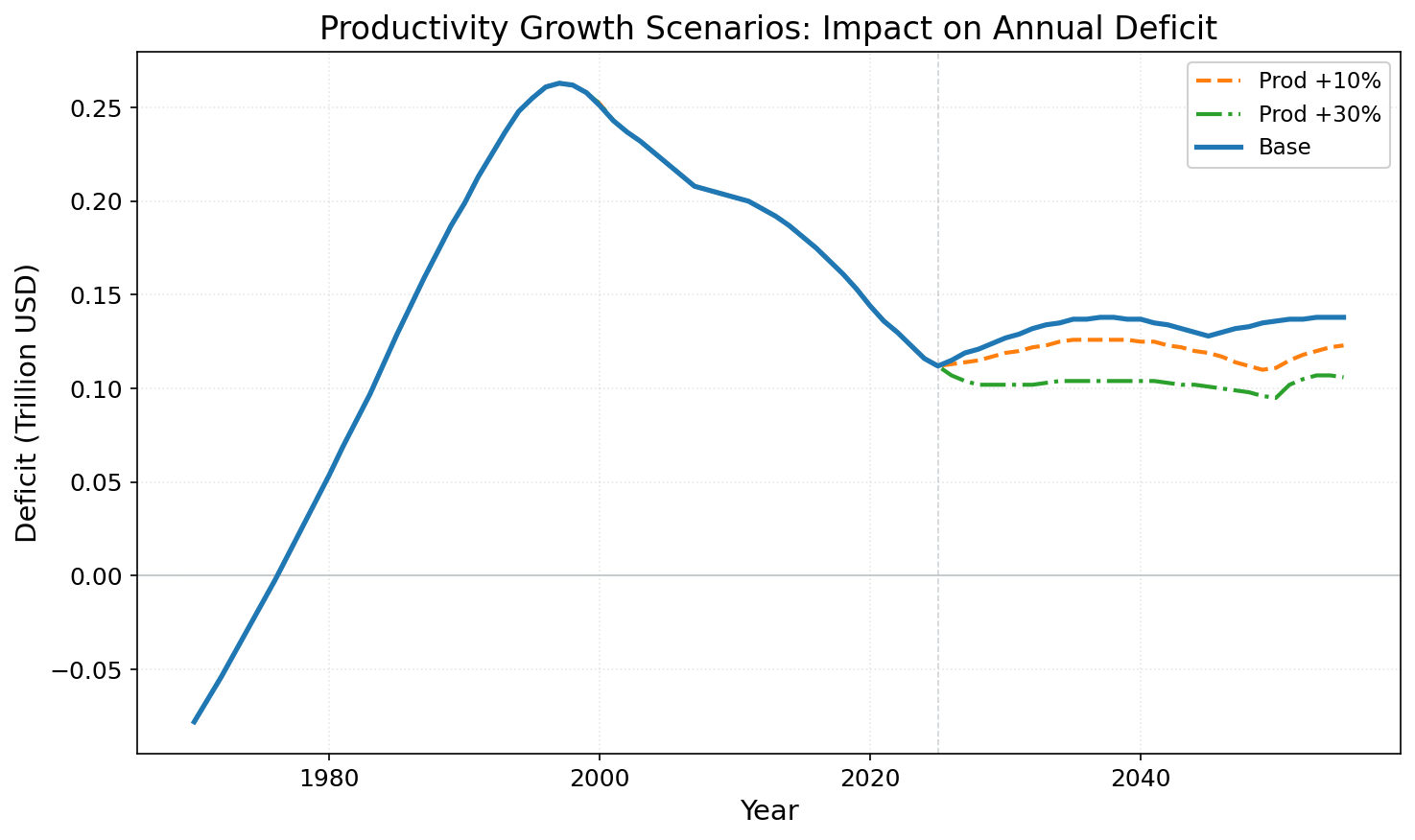}
\caption{Productivity scenarios: baseline vs.\ +10\% and +30\% productivity by 2050 (policy begins in 2025; dashed line).}
\label{fig:policy_productivity}
\end{figure}

Increasing productivity strengthens the revenue-side pathway through GDP. 
In 2050, the moderate productivity scenario reduces the deficit by 17.9\% and net debt by 2.7\% relative to baseline; the aggressive case reduces the deficit by 30.1\% and net debt by 7.1\%. 
Figure~\ref{fig:policy_productivity} shows that productivity improvements act quickly (via GDP and revenue) and can bend the deficit trajectory within a few years of implementation.

\subsection{Scenario 2: Pro-natal policy (fertility increase)}

\begin{figure}[H]
\centering
\includegraphics[width=0.55\linewidth]{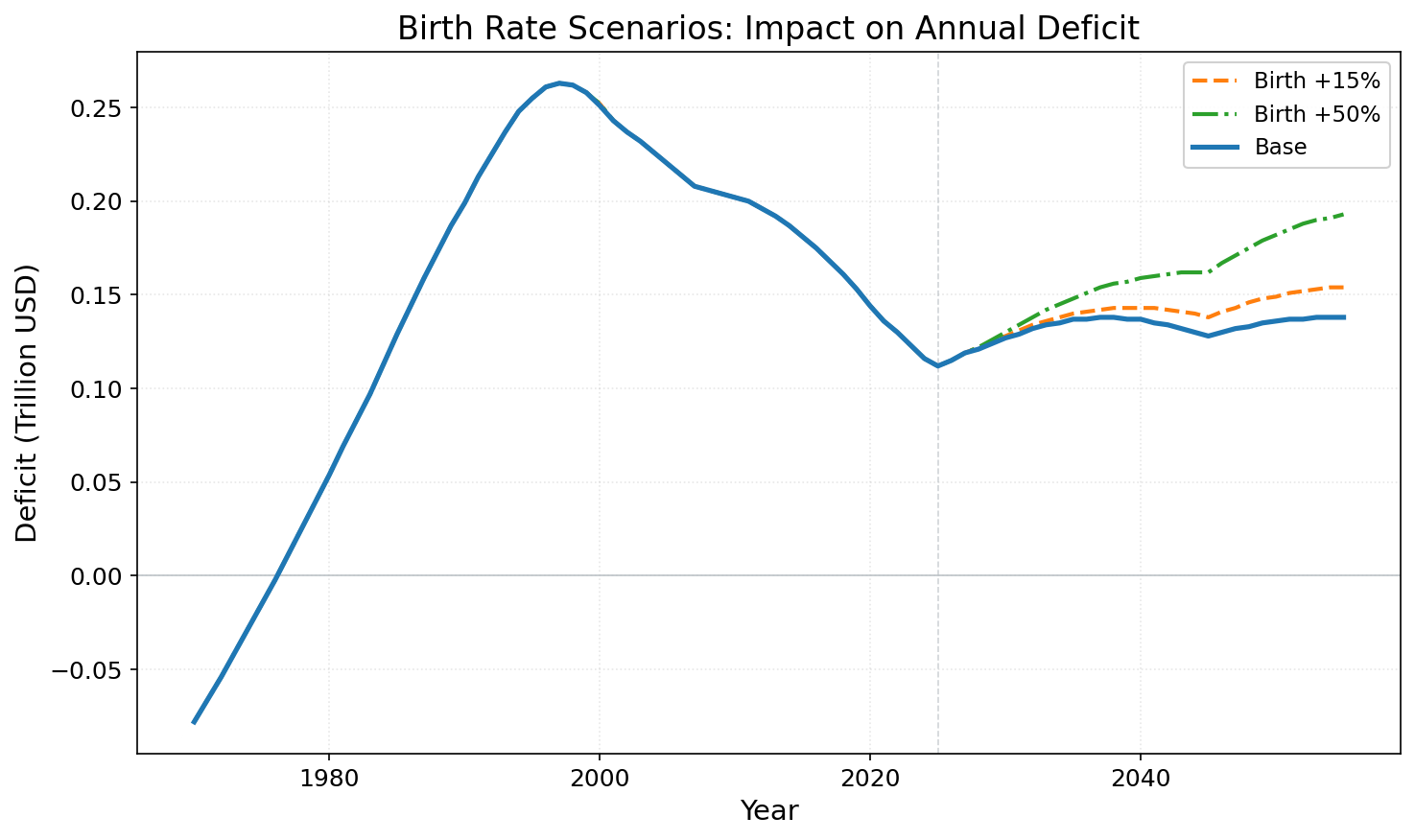}
\caption{Birth-rate parameter scenarios: baseline vs.\ +15\% and +50\% by 2050 (policy begins in 2025; dashed line).}
\label{fig:policy_birth}
\end{figure}

We model births as the product of childbearing-age population and a fractional birth-rate parameter. 
Increasing fertility raises births, but fiscal benefits are delayed because newborn cohorts do not enter the workforce for roughly two decades, while child-related dependency rises immediately. 
This means that higher fertility immediately increases the dependent population and therefore raises spending, even though part of the new cohort becomes productive only after long aging delays.
Consistent with this delay structure, fertility increases worsen the deficit by 2050 in this model: +10.0\% (moderate) and +34.1\% (aggressive). 
Figure~\ref{fig:policy_birth} illustrates the short-run deterioration.

\begin{figure}[H]
\centering
\includegraphics[width=0.55\linewidth]{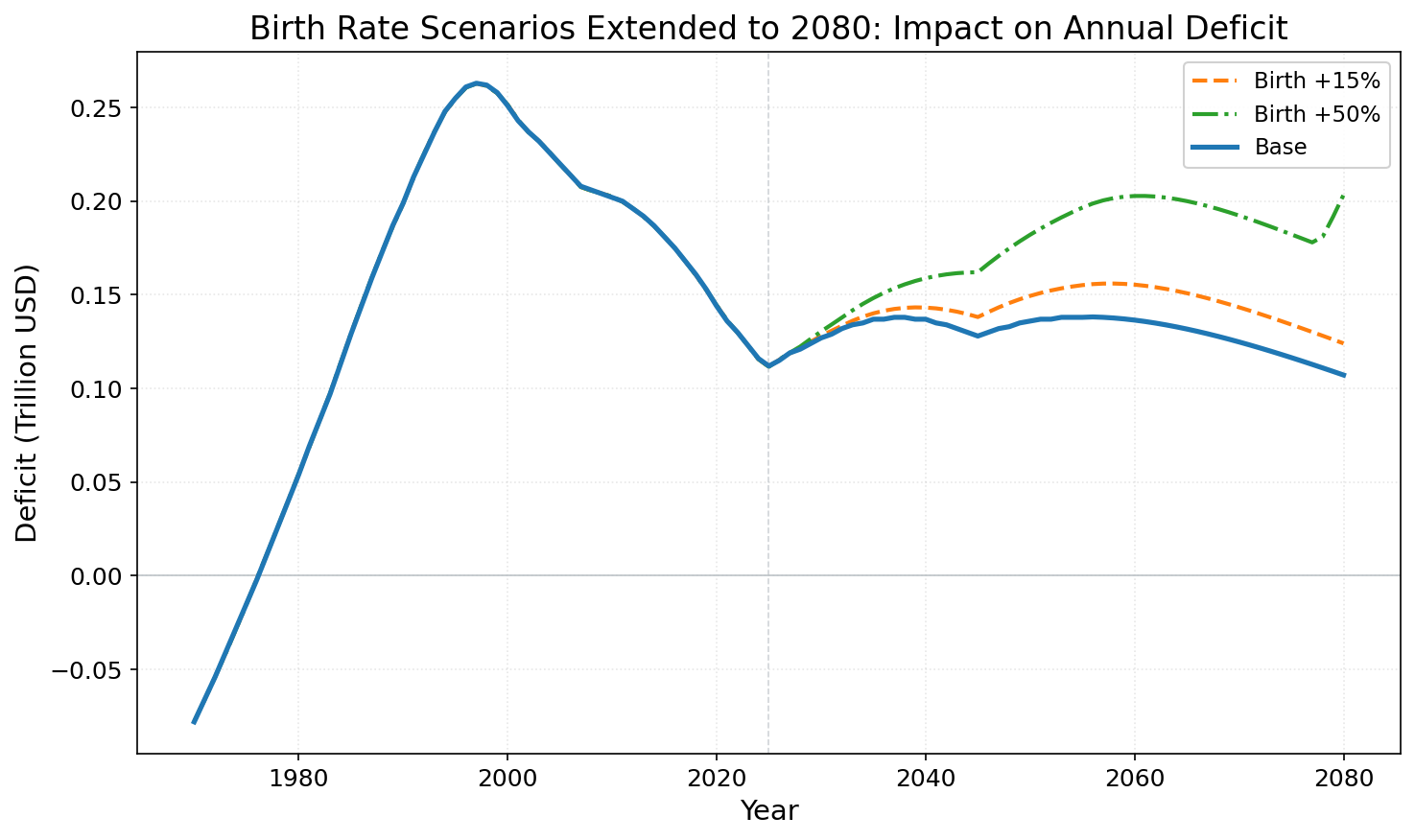}
\caption{Birth-rate parameter scenarios extended to 2080. Deficit remains higher than baseline under the model's aggregated dependent-cost representation.}
\label{fig:policy_birth_long}
\end{figure}

To make the delay implications explicit, Figure~\ref{fig:policy_birth_long} extends the simulation to 2080. Under this aggregated dependent-cost specification, higher fertility does not generate a fiscal ``break-even'' within this timeframe, highlighting the importance of age-specific spending and labor-market assumptions when evaluating pro-natal policies.

\subsection{Scenario 3: Per-capita cost containment}

\begin{figure}[H]
\centering
\includegraphics[width=0.55\linewidth]{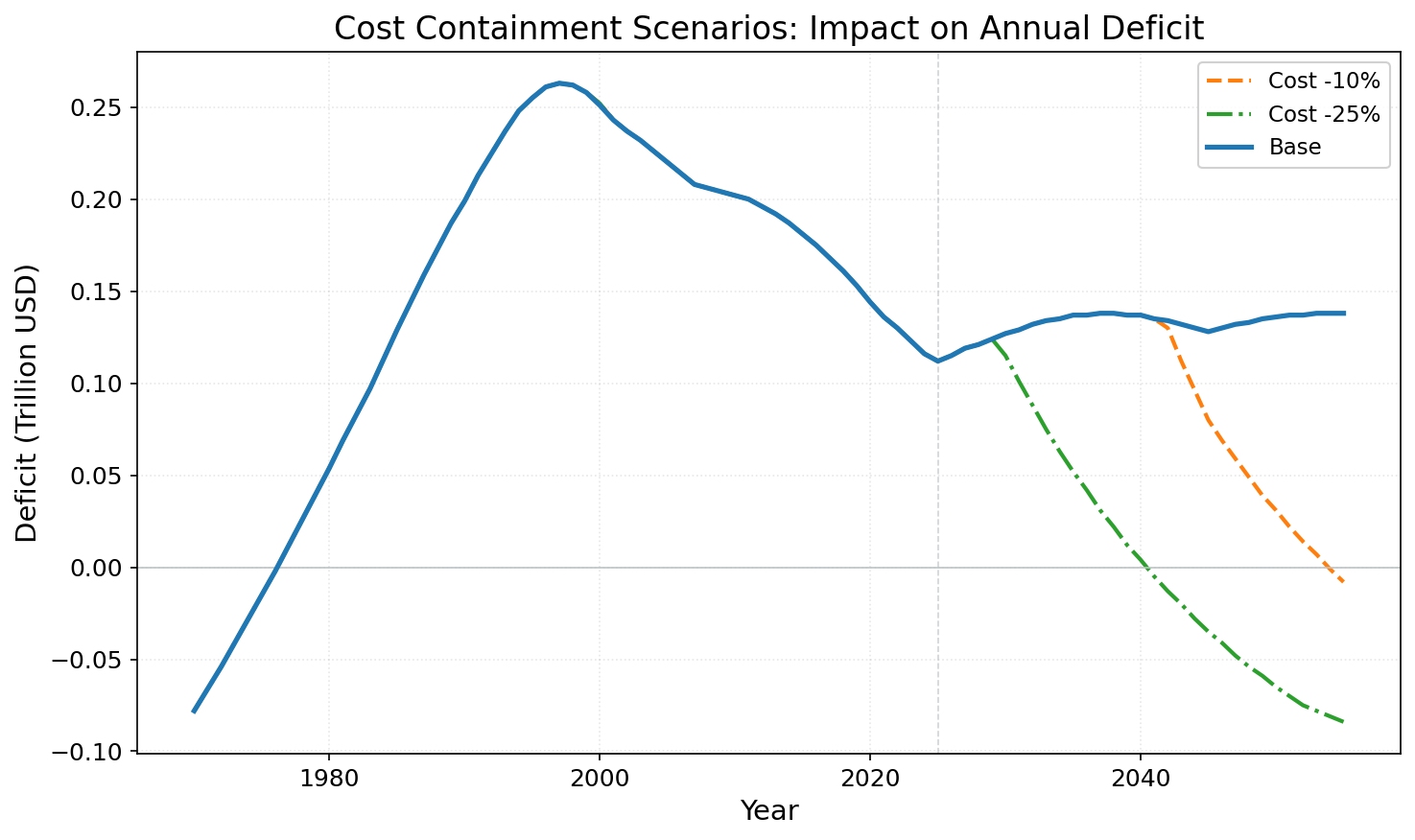}
\caption{Per-capita cost containment scenarios: baseline vs.\ caps that are 10\% and 25\% below the baseline trend by 2050 (policy begins in 2025; dashed line).}
\label{fig:policy_cap}
\end{figure}

Capping (or slowing) per-capita pension/insurance cost directly reduces age-driven spending pressure and interrupts the pathway feeding debt compounding. 
In 2050, the moderate cap reduces the deficit by 77.4\% (substantially closer to balance), and the aggressive cap produces a surplus by 2050 in the model (Figure~\ref{fig:policy_cap}).

\subsection{Combined scenario}

\begin{figure}[H]
\centering
\includegraphics[width=0.6\linewidth]{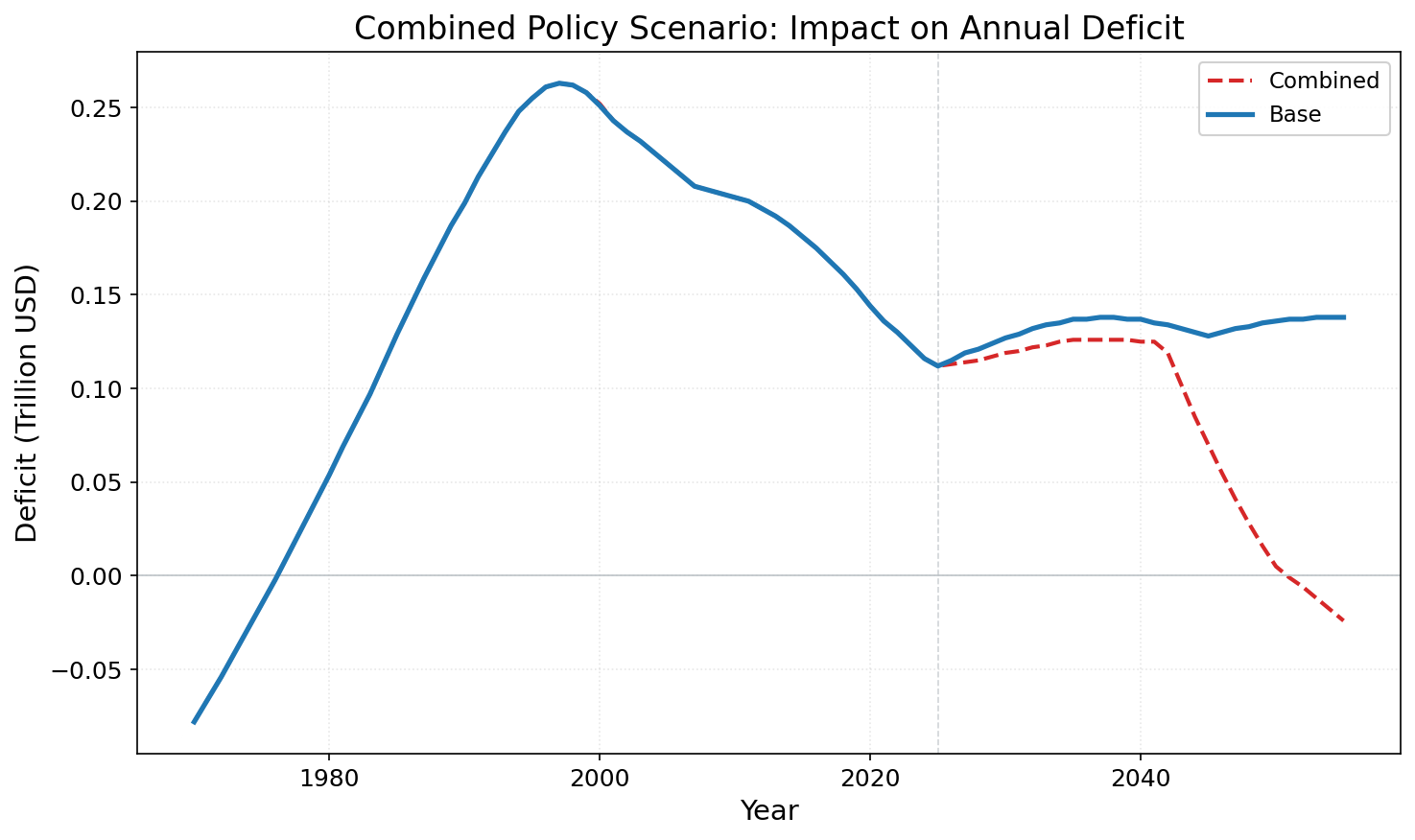}
\caption{Combined moderate scenario (productivity +10\% by 2050 and cost cap 10\% below trend by 2050) vs.\ baseline.}
\label{fig:policy_combined}
\end{figure}

Combining moderate productivity improvement with moderate cost containment yields near-elimination of the 2050 deficit (96.4\% improvement relative to baseline), reflecting complementarity between revenue expansion (GDP) and spending restraint. 
However, the debt stock remains large by 2050 because consolidation occurs gradually and interest costs accumulate over decades (Figure~\ref{fig:policy_combined}).

\section{Discussion}
This study developed and calibrated an integrated system dynamics model of Japan's demographic, economic, and fiscal system to examine fiscal sustainability under demographic pressure. The model connects a seven-stage population aging chain to workforce participation and productivity-driven economic output, which then feeds into a fiscal accounting structure where yearly deficits pile up into debt and interest costs can snowball. We simulate the period from 1968 to 2050 and calibrate the model against historical data on demographics, the economy, and public finances, achieving a reasonable fit ($R^2$ 0.995 for population, 0.989 for GDP, and 0.931 for net debt). We then conducted policy experiments testing three levers: productivity growth, fertility increases, and controlling per-person government costs.

A quantitative macroeconomic literature addresses Japan's aging and fiscal adjustment challenge \citep{imrohoroglu2016, kitao2015, braun2016}. These studies provide rigorous welfare and distributional analyses and quantify the magnitude of adjustment required for fiscal sustainability. Debt-sustainability analyses further emphasize how interest-rate dynamics and fiscal regimes shape projected debt paths \citep{doi2011,hansen2023, hoshi2018}. However, these approaches typically represent policy as simple rules or preset adjustment paths, making it difficult to trace observed fiscal trajectories back to specific feedback structures and implementation delays. Our contribution makes the feedback and delay structure governing deficits, debt, and stabilizing responses explicit, helping explain why policy levers differ sharply in effectiveness within a practical timeframe and why late action becomes disproportionately costly when debt-service dynamics reinforce deficits.

Within the system dynamics tradition, prior work has examined debt sustainability and fiscal adjustment mechanisms across various country contexts \citep{arenas2003,radianti2004,burns2007,john2010,sorci2012}, while Japan-focused SD studies have modeled key subsystems such as healthcare and long-term care costs \citep{inoue2022,nishi2020}. Accounting frameworks have also been developed for Japan \citep{yamaguchi2015asd}. However, these efforts do not focus specifically on feedback under demographic pressure. This paper contributes an end-to-end, data-grounded model that links aging to economic output, revenues, age-related spending, deficits, and debt in one transparent structure. The experiments allow qualitative inference about which feedback pathways are activated by each policy lever: productivity improvements strengthen the link between economic growth and revenue, cost containment directly weakens the pathway from spending to debt, and fertility operates through a delayed demographic channel.

The policy experiments offer practical insights for fiscal policy design. Stabilizing fiscal indicators within decades requires levers that act quickly on the budget—productivity improvements, such as AI implementation and semiconductor manufacturing, and per-capita cost containment---while fertility policies face multi-decade delays before labor-force benefits arrive and can worsen medium-run balances under aggregated dependent-cost assumptions. Because revenue-side and spending-side levers operate through distinct pathways, combining them creates complementary effects: strengthening the stabilizing forces while weakening the interest-cost snowball. The model can serve as a transparent scenario laboratory for policy sequencing questions, and future extensions could incorporate immigration scenarios, endogenous debt-to-rate feedback, and cross-country comparison among aging high-debt economies.

Several limitations should be noted. The model is designed for long-run structure rather than short-run economic fluctuations; it smooths business-cycle fluctuations and does not represent detailed interactions between monetary and fiscal policy. Net-debt fit is weaker than population and GDP ($R^2$ 0.931), reflecting definitional and accounting differences relative to the model's simplified identity in which debt is the integral of deficits. The spending side is intentionally kept simple, so the fertility-policy results depend on simplified assumptions about child-related costs and labor-market responses. Finally, the model does not include political economy constraints that may limit how aggressively governments can cut costs in practice.

Japan's fiscal challenge under demographic aging is fundamentally a problem of accumulation, delay, and feedback. By making these mechanisms explicit, our model clarifies which stabilizing forces matter most within a policy-relevant timeframe. These insights support the design of time-aware policy packages and provide a platform for more detailed analyses of aging, high-debt economies.

\clearpage
\appendix

\section{Appendix: Data sources}
This appendix documents the primary data sources used to construct the reference modes, calibration targets, and figure data.
\begin{small}
\begin{longtable}{@{} p{0.28\linewidth} p{0.42\linewidth} p{0.24\linewidth} @{}}
\caption{Primary data sources for reference modes, calibration targets, and unit conversions.}\label{tab:data_sources}\\
\toprule
\textbf{Series / model input} & \textbf{Parameters} & \textbf{Provider}\\
\midrule
\endfirsthead
\toprule
\textbf{Series / model input} & \textbf{Parameters} & \textbf{Provider}\\
\midrule
\endhead
\midrule \multicolumn{3}{r@{}}{\textit{(continued)}}\\
\endfoot
\bottomrule
\endlastfoot
\multicolumn{3}{@{}l}{\textbf{Demographic series}}\\
\addlinespace[3pt]
Population & Population by age group (0--14, 15--24, 25--34, 35--44, 45--54, 55--64, 65+); Population & e-Stat \citep{estat1960-2000,estat2000-2020}\\
\addlinespace[3pt]
Birth rate & Birth rate; fractional birth rate of U45 & e-Stat \citep{estat_birth}\\
\addlinespace[3pt]
Death rate & Young death rate; Elderly Death rate & IPSS \citep{ipss_jmd2024}\\
\addlinespace[3pt]
\multicolumn{3}{@{}l}{\textbf{Labor market series}}\\
\addlinespace[3pt]
Workforce & Middle-age workforce; elderly workforce; total workforce; participation rates by age & e-Stat \citep{estat_labor}\\
\addlinespace[3pt]
Average productivity & Average productivity (GDP/workforce) & Calibrated\\
\addlinespace[6pt]
\multicolumn{3}{@{}l}{\textbf{Economic series}}\\
\addlinespace[3pt]
GDP & GDP (constant 2015 US\$) & World Bank \citep{wb_gdp_jp}\\
\addlinespace[3pt]
GDP per capita & GDP per capita (constant 2015 US\$) & World Bank \citep{wb_gdp_per_capita_jp}\\
\addlinespace[6pt]
\multicolumn{3}{@{}l}{\textbf{Fiscal series}}\\
\addlinespace[3pt]
Fiscal aggregates & Government revenue; pension/insurance cost; interest on debt; other cost; government spending; deficit; gross debt; net debt; other revenue & Cabinet Office \citep{cao_gfs}\\
\addlinespace[3pt]
Tax rate & Effective tax rate (national burden ratio) & MOF Japan \citep{mof_burden}\\
\addlinespace[6pt]
\end{longtable}
\end{small}

\clearpage
\printbibliography

\end{document}